\documentclass[prl,twocolumn,superscriptaddress,preprintnumbers,nofootinbib,APS]{revtex4-1}
\usepackage{color}
\usepackage{psfrag}
\usepackage{dcolumn}
\usepackage{bm}
\usepackage[latin1]{inputenc}
\usepackage[spanish,english]{babel}
\usepackage{amsfonts}
\usepackage{amssymb,epsf}
\usepackage{graphicx}
\usepackage{epstopdf}
\usepackage{epsfig}
\usepackage{latexsym}
\usepackage{epsf}

\usepackage{mathtools,mathptmx,array,slashed} 

\usepackage{graphicx,graphics,subfigure}
\usepackage{ctable,multirow}
\usepackage{tabularx}
\usepackage{caption}

\begin{document}

\title{\textbf{
On the Classical, Penrose, and Reverse Isoperimetric Inequalities in Black Holes:\\Insights From AdS to Flat Riemannian Backgrounds}}


\author{Robert B. Mann}
\email{rbmann@uwaterloo.ca}
\affiliation{{\footnotesize Department of Applied Mathematics, University of Waterloo, Waterloo, Ontario N2L 3G1, Canada}}
\affiliation{{\footnotesize Institute for Quantum Computing, University of Waterloo, Waterloo, Ontario N2L 3G1, Canada}}
\affiliation{{\footnotesize Perimeter Institute for Theoretical Physics, 31 Caroline St. N., Waterloo, Ontario N2L 2Y5, Canada}}
\affiliation{{\footnotesize Department of Physics and Astronomy, University of Waterloo, Waterloo, Ontario N2L 3G1, Canadaa}}
\author{Behnam Pourhassan}
\email{b.pourhassan@du.ac.ir}
\affiliation{{\footnotesize School of Physics, Damghan University, Damghan 3671641167, Iran}}
\affiliation{{\footnotesize Center for Theoretical Physics, Khazar University, Baku, Azerbaijan}}
\author{Ali Dehghani}
\email{ali.dehghani.phys@gmail.com}
\affiliation{{\footnotesize School of Physics, Damghan University, Damghan 3671641167, Iran}}

\begin{abstract}
A collection of evidence is presented showing that the conjectured reverse isoperimetric inequality (RII) for asymptotically anti-de Sitter (AdS) as well as de Sitter (dS) black holes, ${\cal R}_\text{RII} \ge 1$, is of fundamental importance and its violation (${\cal R}_\text{RII} < 1$) gives rise to either thermodynamic instabilities, or physically unreasonable solutions, or naked singularities. We show that AdS black holes violating reverse isoperimetric inequality, known as superentropic black holes, satisfy neither mechanical stability requirement of ${\kappa _T} \geqslant {\kappa _S} \geqslant 0$ nor thermal stability requirement of ${C_P} \geqslant {C_V} \geqslant 0$. This property makes them thermodynamically unstable for the whole range of parameter space and explains all the diverse behaviors so far reported. We conjecture this statement holds for all superentropic black holes. We confirm that there is no counterexample to the thermodynamic instability conjecture of superentropic black holes and extend this to dS space by presenting the first example of superentropicity in dS space. By bringing evidence, we then propose two other conjectures: 1) there is no asymptotically flat limit of superentropic black holes, and 2) there exist two stronger versions of Penrose isoperimetric inequality (PII) for asymptotically flat black holes; the first is the asymptotically flat limit of the RII and the second is a new inequality in terms of the ADM mass and the thermodynamics volume, that we call thermo-volumetric inequality, obtained via insights from the $\Lambda \to 0$ limit of extended black hole thermodynamics in AdS. We show that the Penrose isoperimetric inequality (PII) is weaker than the reverse isoperimetric inequality (RII),
as demonstrated explicitly for $D=4,5$ and for all black hole families we have examined.

\end{abstract}

\maketitle

\section{Classical, Penrose, and reverse isoperimetric inequalities}

The robust prediction of black holes \cite{Penrose1965,Penrose1969,Thorne1970,RuffiniWheeler1971} by Einstein's theory of General Relativity \cite{Eistein1915,Einstein1916} as well as its modifications \cite{ModifiedGR2015} is one of the most fascinating and stunning achievements in theoretical physics, which is now widely based on empirical evidence \cite{Genzel2010,Ghez2008,LIGO2016,LIGO2017Virgo,Gravity2018,EHT2019}. These mysterious objects provide a basis for studying quantum effects in gravity \cite{Mukhanov2007Book}  potentially opening a path to quantum gravity via their thermodynamic properties \cite{Wald1999,Carlip2014}. Black holes obey four laws that are very similar to those of classical thermodynamics \cite{BCH1973}, with some differences, e.g., the relativistic mass ($M c^2$) replaces the usual concept of energy in classical physics and the corresponding temperature has a quantum mechanical origin that is proportional to the surface gravity of the black hole \cite{Hawking1975}. Black hole thermodynamics is strongly connected to the \textit{shape} (geometry) of the black hole; the entropy is proportional to the area of the event horizon ($A$),  a global property, and thermodynamic quantities  satisfy several important geometric inequalities \cite{Gibbons2012,Dain2014,Dain2018LRR}. 

For asymptotically flat black holes with charge and rotation parameters  the semi-classical first law and   corresponding Euler relation (better known as the Smarr formula \cite{Smarr1973}) are respectively (in geometrical units $G_N = c = k_B = \hbar = 1$)
$dM=TdS + \Omega dJ + \Phi dQ$ and $M = 2TS+ 2 \Omega J + \Phi Q$, where $S$, $\Omega$, $J$, $\Phi$, and $Q$ stand for the respective entropy, angular velocity, angular momentum, electric potential, and charge \cite{BCH1973}.  A remarkable  connection between the shape and thermodynamics of asymptotically flat black holes is the Penrose Isoperimetric Inequality (PII)\footnote{While the original formulation by Penrose is known simply as the Penrose inequality, we refer to it here as the Penrose Isoperimetric Inequality to emphasize its structural analogy with other geometric inequalities of isoperimetric type. In this context, the horizon area plays the role of an effective perimeter bounding the ADM mass, just as in the classical isoperimetric inequality the boundary length bounds the enclosed area. This terminology highlights the unified geometric character of the classical, Penrose, and reverse isoperimetric inequalities considered in this work.}
in four-dimensions \cite{Penrose1973}, which states that the area $A$ of the event horizon is bounded above by the ADM mass (that is the total relativistic energy, $E=M$), i.e. $A \le 16 \pi M^2$,   under the assumption of gravitational collapse together with the cosmic censorship hypothesis~\cite{Penrose1969}, which ensures that the end state of collapse is a black hole rather than a naked singularity.  Additionally, some generalizations to higher dimensions suggest the following result \cite{Bray2001,Ida2002,Holzegel2006,Bray2007}
\begin{equation} \label{Penrose inequality}
{\left( {\frac{A}{{{{\cal A}_{D - 2}}}}} \right)^{D - 3}} \le {\left( {\frac{{16\pi M}}{{(D - 2){{\cal A}_{D - 2}}}}} \right)^{D - 2}},
\end{equation}
where ${\cal A}_{D-2}$ is the volume of a unit $(D-2)$-sphere. Extension to spacetimes with general horizons (not only the outermost minimal surfaces) in asymptotically flat Riemannian 3-manifolds are also discussed in \cite{Malec2002}. The Riemannian PII (\ref{Penrose inequality}) is an important example of an isoperimetric inequality in physics, playing a vital role in the cosmic censorship hypothesis;  its violation implies naked singularities. 

In mathematics, the simplest situation for illustrating the isoperimetric inequality is plane surfaces in two dimensional Euclidean space $\mathbb{R}^2$: the area $A$ of any plane figure satisfies the inequality $4 \pi A \le \ell^2$, in which $L$ is the length of a closed curve that encloses $A$ and equality holds if and only if the curve is a circle. This statement  generalizes to higher-dimensional Euclidean spaces $\mathbb{E}^{D-1}$ for the perimeter (surface area) of a connected set ($A$) and its volume ($V$), which is mathematically stated as
\begin{equation} \label{classicalII}
{\left( {\frac{{(D - 1)V}}{{{{\cal A}_{D - 2}}}}} \right)^{D - 2}} \le {\left( {\frac{A}{{{{\cal A}_{D - 2}}}}} \right)^{D - 1}},
\end{equation}
or alternatively we can write this as the Conventional Isoperimetric Inequality (CII),
\begin{equation} \label{II_Euclidean}
{\cal R}_\text{CII} \equiv {\left( {\frac{{(D - 1)V}}{{{{\cal A}_{D - 2}}}}} \right)^{\frac{1}{{D - 1}}}}{\left( {\frac{{{{\cal A}_{D - 2}}}}{A}} \right)^{\frac{1}{{D - 2}}}} \le 1
\end{equation}
where ${\cal R}_\text{CII}$ is the (dimensionless)  isoperimetric ratio and  equality holds only when the object is a (hyper)sphere.  

At first glance, it seems unlikely to find a connection between the PII (\ref{Penrose inequality}) and the CII (\ref{II_Euclidean}). This is because   the former combines a thermodynamic quantity (the relativistic mass, $M$) with a geometric one (the closed trapped surface $A$), but in the latter all quantities are purely geometric without closed trapped surfaces. Furthermore, there is no widely accepted definition for the volume of a black hole \cite{Review2017CQG}.

However in the last two decades thermodynamic concepts of volume and pressure, absent in the original formulation of black hole thermodynamics \cite{BCH1973}, have been introduced. The negative cosmological constant ($\Lambda < 0$) is regarded as thermodynamic pressure \cite{Kastor2009} (which also extends to $\Lambda > 0$),  which leads to a natural definition of a conjugate volume as well as the emergence of a number of novel physical phenomena in black holes with internal consistency in a way reminiscent of chemical thermodynamics \cite{Page2019,Review2017CQG,Mann:2025xrb}. This framework, known black hole chemistry, is derived through generalizing Komar's method\footnote{Although the generalized Komar integral and its relation to the thermodynamic volume have been discussed in several of the cited works, we briefly include it here to make	the presentation self-contained and to clarify how it naturally leads to the extended first law and Smarr relation used in our subsequent analysis. 
This explicit reminder ensures consistency of notation and emphasizes the origin of 
the thermodynamic volume $V$ employed throughout this paper.}  \cite{Komar1959} to include the cosmological constant as \cite{Zyla, Kastor2009} 
\begin{equation} \label{KomarInt}
\frac{{d - 2}}{{8\pi {G_N}}}\oint_{\partial \Sigma } {d{S_{\mu \nu }}\left( {{\nabla ^\mu }{\xi ^\nu } + \frac{2}{{d - 2}}\Lambda {\omega ^{\mu \nu }}} \right)}  = 0,
\end{equation}
where $d{S_{ab}}$ is the volume element normal to the co-dimension 2 surface $\partial \Sigma $, and the Killing potential ${\omega ^{\mu\nu}}$ is defined from the Killing vector  ${\xi ^\mu} = {\nabla _\nu}{\omega^{\mu\nu}}$. 
This generalized Komar integral follows directly from Einstein's equations with a cosmological constant in the absence of matter fields. 
In particular, the right-hand side vanishes because we are assuming vacuum Einstein equations 
$R_{\mu\nu} - \tfrac{1}{2}R g_{\mu\nu} + \Lambda g_{\mu\nu} = 0$.
If additional stress-energy sources are present, one obtains an extra contribution 
$\sim \int_{\Sigma} T_{\mu\nu}\xi^{\nu} d\Sigma^{\mu}$, so that the integral in (\ref{KomarInt}) would not vanish in general. It turns out that the extended first law and the corresponding Smarr relation (in agreement with scaling argument) are respectively obtained as
\begin{gather} 
dM = TdS + VdP + ... \, \label{extended_first_law},\\ 
(D - 3)M = (D - 2)TS - 2PV + ... \, \label{extended_Smarr},
\end{gather}
in $D$ spacetime dimensions, 
where the ellipsis accounts for other possible work terms, and \cite{Kastor2009,Review2017CQG,Mann:2025xrb}
\begin{gather}\label{press}
P=\frac{\Lambda}{8 \pi G_N}=-\frac{(D-1)(D-2)}{16 \pi G_N \ell^2}, \\ \quad V={\int_{\partial {\Sigma _\infty }} {d{S_{\mu\nu}}\left( {{\omega ^{\mu\nu}} - \omega _{{\rm{AdS}}}^{\mu\nu}} \right) - \int_{\partial {\Sigma _{\rm{h}}}} {d{S_{\mu\nu}}{\omega ^{\mu\nu}}} } }
\label{volume-thermo}
\end{gather}
where  $\omega _{\rm{AdS}}^{\mu\nu}$ is the Killing potential for AdS spacetime. 
The quantity \(V\),  interpreted strictly as the thermodynamic volume, is conjugate to the pressure $P=-\Lambda/(8\pi G_N)$ in the extended first law,
\begin{equation} \label{Volconj}
V \equiv \left(\frac{\partial M}{\partial P}\right)_{S,J,Q,\ldots},
\end{equation}
with $M$ regarded as the enthalpy of the spacetime~\cite{Kastor2009,Cvetic2011}. This definition is covariant (it can be written via the Killing potential as in (\ref{volume-thermo})), and is the unique choice that simultaneously satisfies the first law (\ref{extended_first_law}) and the Smarr relation (\ref{extended_Smarr}). It should not be conflated with the proper geometric volume inside the horizon, which is slicing dependent and generally inequivalent to $V$. For static, spherically symmetric solutions one finds $V=\Omega_{D-2} r_+^{D-1}/(D-1) = V_{\text{geo}}$, coinciding with the naive geometric volume, whereas for rotating solutions $V\neq V_{\text{geo}}$, reflecting the contribution of rotation to the enthalpy at fixed $S,J,\ldots$. 

A striking result   that emerged from  black hole chemistry is the conjectured Reverse Isoperimetric Inequality (RII), which states that anti-de Sitter (AdS) black holes satisfy a geometric inequality that is precisely the reverse of the  Euclidean CII  (\ref{II_Euclidean}). The RII conjecture is mathematically formulated as \cite{Cvetic2011},
\begin{equation} \label{RII}
{\cal R}_\text{RII} \equiv {\left( {\frac{{(D - 1)V}}{{{{\cal A} _{D - 2}}}}} \right)^{\frac{1}{{D - 1}}}}{\left( {\frac{{{{\cal A} _{D - 2}}}}{A}} \right)^{\frac{1}{{D - 2}}}} \ge 1,
\end{equation}
where $V$ is the thermodynamic volume (\ref{volume-thermo}) of the black hole, $A$ is the area of the outer (event) horizon, and ${\cal A}_{D-2}$ the corresponding area of the unit $(D-2)$-hypersurface,  including sphere, plane, and hyperbolic space. The RII has been also conjectured to be valid for de Sitter black holes for the thermodynamic volumes associated with the event and cosmological horizons \cite{Dolan2013deSitterIso}. The RII (\ref{RII}) implies that a given volume $V$ encloses the maximal possible area when the inequality is saturated, ${\cal R}_\text{RII} =1$. Alternatively, the volume is bounded from below when the area $A$ is held fixed. Since for black holes the area of the outer horizon is linearly proportional to the entropy $S$, the RII conjecture can be rephrased as: the entropy inside a horizon of a given volume $V$ is maximized when ${\cal R}_\text{RII}=1$.

While this elegant conjecture has been demonstrated to hold true for numerous families of AdS black holes, there have also been instances where it has been violated, through the so-called superentropic 
black holes \cite{UltraSpinningBH2015PRL}, for which the associated entropy exceeds the maximal bound implied by the RII conjecture. Examples are charged BTZ black holes \cite{Johnson2019} and their higher-dimensional generalizations \cite{DPZS2023} as well as nonlinear $U(1)$ modifications \cite{CE2021BIBTZ}, different classes of ultraspinning superentropic black holes \cite{UltraSpinningBH2015PRL,UltraSpinningBH2015JHEP}, generalized exotic BTZ black holes \cite{Cong2019Mann}, superentropic black holes with Immirzi hair \cite{Immirzi2021} and more. Later, by analyzing specific heats of superentropic black holes, it was conjectured that superentropicity (${\cal R}_\text{RII} < 1$) is somehow connected to a new, fundamental thermodynamic instability \cite{Johnson2019,Cong2019Mann}, which gives the RII conjecture (\ref{RII}) a special significance.

There are some claimed counterexamples \cite{CE2021BIBTZ,CE2023KerrNewmanAdS,CE2023TorusBH,CE2023CG} that challenge the instability conjecture of superentropic black holes and (if true) subsequently rule out the fundamentality of the RII conjecture, but their validity is questionable to us. Further investigation is needed to determine whether these counterexamples are truly valid or not, as will be addressed fully here. On the other hand, exploring the interrelation between the aforementioned three isoperimetric inequalities (\ref{Penrose inequality}), (\ref{II_Euclidean}), and (\ref{RII}) poses a challenge due to their formulation in distinct spaces. Can these three be linked in a same background through a sort of limiting procedure? If possible, this could result in new isoperimetric inequalities for black holes and advance our understanding of their potential applications and implications.

\section{Our goals} 

We are  interested in the physical significance and implications of the  RII conjecture  \eqref{RII}. 
Specifically we shall address the following questions.  (i) What are the thermodynamic implications of violating the RII (\ref{RII})? (ii) What are the  asymptotically flat limits of (A)dS black holes with ${\cal R}_\text{RII} \ge 1$ and ${\cal R}_\text{RII} < 1$?  (iii) What, if any, is the relationship  between the RII (formulated in asymptotically AdS spacetimes) and the PII (formulated in asymptotically flat spacetimes)?

We find that violating the conjectured RII corresponds to violating  both the mechanical stability requirements of positive 
adiabatic and isothermal compressibilities (respectively ${\kappa _T} \geqslant {\kappa _S} \geqslant 0$)  and the themal stability requirements
of positive specific heat at constant pressure and constant volume 
(respectively ${C_P} \geqslant {C_V} \geqslant 0$)  for physical systems. We emphasize that there is no counterexample. An important observation is that superentropic black holes also suffer from negative compressibility. Using the results of  extended black hole thermodynamics in AdS and taking the limit $\Lambda \to 0$, we then present a set of isoperimetric inequalities for four-dimensional flat Riemannian backgrounds that  are conjectured to remain valid for all asymptotically flat black holes. We also argue that an asymptotically flat limit of superentropic black holes does not exist, otherwise the   PII would be violated, resulting in a naked singularity. By taking the flat limit of the generalized Komar integral \eqref{KomarInt} in AdS, we propose a new isoperimetric inequality purely in terms of the thermodynamic quantities of $M$ and $V$, that we call thermo-volumetric inequality. We also speculate that a violation of the thermo-volumetric inequality could be related to a new kind of black hole instability.

\section{Superentropicity implies thermodynamic instabilities}  

It will be constructive to first understand sub/super-entropicity, i.e. respecting/violating the RII conjecture and its thermodynamical implications, intuitively. This is helpful to find the appropriate way of revealing the potential thermodynamic instabilities in superentropic black holes. The word \textit{isoperimetric} mathematically means \textit{having the same perimeter} (here $A$) which for black holes translates to \textit{having the same entropy}. If  ${\cal R}_\text{RII} > 1$ then  the volume cannot decrease indefinitely for a fixed amount of entropy,   consistent with our physical intuition. The respective adiabatic and isothermal compressibilities 
\begin{equation}\label{11}
{\kappa _{S}} =  - \frac{1}{V}{{\left. {\frac{{\partial V}}{{\partial P}}} \right|}_{S}} \, , \quad {\kappa _{T}} =  - \frac{1}{V}{{\left. {\frac{{\partial V}}{{\partial P}}} \right|}_{T}},
\end{equation}
measure the change in volume   of the system in  response to   compression (a change in pressure). If $\kappa_S>0$ then within the physically stable region of parameter space the system expands under a decrease in pressure during an adiabatic process.  When AdS black holes respect 
but do not saturate  the inequality \eqref{RII}, they can change their size during adiabatic processes  
whilst maintaining positive adiabatic compressibility ($\kappa_S > 0$) for certain range of the parameter space. Well-known examples are the Kerr(-Newman)-AdS class of black hole spacetimes, which have positive $\kappa_{S}$ \cite{Dolan2011},  always respect the RII conjecture (${\cal R}>1$) \cite{Cvetic2011}, and in the non-rotating ($a \to 0$) limit saturate the inequality (${\cal R}_\text{RII}=1$), attaining $\kappa_{S}=0$. 
In other regions of parameter space, however, $\kappa_S$ can indeed become negative, signalling mechanical instability. Such cases will be discussed later in the context of superentropic black holes where $\kappa_S<0$.

Next, consider AdS black holes that always saturate the inequality, ${\cal R}_\text{RII}=1$. In such cases, the black hole acquires the maximum entropy allowed according to the conjecture. If the area $A$ is kept fixed, which corresponds to fixing the entropy $S$, then saturation implies no change in volume, $\delta V =0$. In short, if ${\cal R}_\text{RII}=1$ everywhere then $\left. \delta V \right|_S =0$. This results in incompressibility of the black hole system during adiabatic processes, $\kappa_S=0$. With the same logic, starting from ${\cal R}_\text{RII}=1$ together with the isochoric assumption $\delta V =0$ results in $\left. \delta S \right|_V =0$, meaning that the specific heat at constant volume must be zero, $C_V=0$. While the results of $\kappa_S=0$ and $C_V=0$ may be familiar for many black hole spacetimes  (mostly static ones) in AdS, it is worth noting that they were derived from the RII conjecture, which serves as our starting point. This adds an interesting perspective to the fundamentality of the isoperimetric quotient ${\cal R}_\text{RII}$.

One should now ask what the thermodynamic implications of violating ${\cal R}_\text{RII} \ge 1$ are in black holes, particularly when it is subjected to changes in pressure, while keeping  either its temperature or entropy constant. Note that the violation of the conjectured RII (\ref{RII}) necessitates the absence of a non-trivial lower bound, yet we now have an upper bound instead. Surprisingly, we generally observe $\Delta V > 0$ while $\Delta P > 0$, meaning that by constantly lowering the pressure, the effective volume shrinks 
and vice versa (as will be shown explicitly in the examples below, where $(\partial V / \partial P)_{S,T} > 0$ for all superentropic cases).
From this, a tendency to attain negative compressibility is understandable. This suggests that violating ${\cal R}_\text{RII} \ge 1$ has a profound effect on the compressibility as well as the mechanical and thermal stability of the system.

So the first clue to reveal the relevant thermodynamic instabilities is to begin checking compressibilities. No matter what theory we are dealing with, all the superentropic black holes in the extended phase space certainly have the four quantities $T$, $S$, $V$, and $P$. In our considerations, we only deal with these, and suppress the rest of thermodynamic quantities for simplicity of notation. 

Let us first take a look at the stability requirements for a thermodynamic system that can be established in terms of the concavity of entropy, the convexity of internal energy, or extending these conditions for any possible Legandre
transforms of energy such as Helmholtz free energy, Gibbs energy, enthalpy etc. In summary, in the energy representation we have:
\begin{equation}
{E_{SS}}: = \underbrace {{{\left. {\frac{{{\partial ^2}E}}{{\partial {S^2}}}} \right|}_V}}_{ = T/{C_V}} \ge 0 \quad \text{and} \quad {E_{VV}}: = \underbrace {{{\left. {\frac{{{\partial ^2}E}}{{\partial {V^2}}}} \right|}_S}}_{ = {{({\kappa _S}V)}^{ - 1}}} \ge 0,
\end{equation}
while, in the enthalpy ($H \equiv M=E+PV$) representation: 
\begin{equation}
{H_{SS}}: = \underbrace {{{\left. {\frac{{{\partial ^2}H}}{{\partial {S^2}}}} \right|}_P}}_{ = T/{C_P}} \ge 0 \quad \text{and} \quad {H_{PP}}: = \underbrace {{{\left. {\frac{{{\partial ^2}H}}{{\partial {P^2}}}} \right|}_S}}_{ =  - V{\kappa _S}} \le 0,
\end{equation}
and, in Gibbs free energy ($G=H-TS$) representation: 
\begin{equation}
{G_{TT}}: = \underbrace {{{\left. {\frac{{{\partial ^2}G}}{{\partial {T^2}}}} \right|}_P}}_{ =  - {C_P}/T} \le 0 \quad \text{and} \quad {G_{PP}}: = \underbrace {{{\left. {\frac{{{\partial ^2}G}}{{\partial {P^2}}}} \right|}_T}}_{ =  - V{\kappa _T}} \le 0.
\end{equation}
$\kappa_{T}$ and $\kappa_{S}$ are isothermal and adiabatic compressibilities and, ${C_P} = T{\left. {(\partial S/\partial T)} \right|_P}$ and ${C_V} = T{\left. {(\partial S/\partial T)} \right|_V}$ are isobaric and isochoric specific heats, respectively. In a thermodynamically stable system, both specific heat capacities as well as both compressibilities must be positive, satisfying 
\begin{equation} \label{thermo identities}
{\kappa _T} - {\kappa _S} = \frac{{TV\beta _P^2}}{{{C_P}}} \qquad {\rm{and}} \qquad {C_P} - {C_V} = \frac{{TV\beta _P^2}}{{{\kappa _T}}},
\end{equation}
where ${\beta _P} = (1/V){\left. {(\partial V/\partial T)} \right|_P}$ is the isobaric expansivity. The above relations imply the following mechanical and thermal stability requirements
\begin{equation} \label{stability criteria}
{{\kappa _T} \ge {\kappa _S} \ge 0} \qquad \text{and}   \qquad   {{C_P} \ge {C_V} \ge 0},
\end{equation}
respectively. From these, the  adiabatic index, $\gamma$, is  defined to be
\begin{equation} \label{adiabatic index}
\gamma  \equiv \frac{{{C_P}}}{{{C_V}}} = \frac{{{\kappa _T}}}{{{\kappa _S}}},
\end{equation}
from which the equality between the ratio of compressibilities and that of specific heats is simply proved using eqs. (\ref{thermo identities}). The stability requirements (\ref{stability criteria}) imply that the adiabatic index must be greater than or equal to unity ($\gamma \ge 1$), as occurs  in nature.  Furthermore, to be more rigorous, we have directly computed all the thermodynamic coefficients to make sure the adiabatic index relation (\ref{adiabatic index}) holds in all the cases, which is also a confirmation of the thermodynamic identities in (\ref{thermo identities}), as expected.

In what follows, we   explore the  thermodynamic implications of violating RII   by going through a couple of examples. We explicitly show that superentropicity in black holes is connected to thermodynamic instabilities; they all suffer from negative compressibility and from (\ref{thermo identities}) and (\ref{adiabatic index}),  it becomes evident that they cannot satisfy the thermodynamic stability requirements in   (\ref{stability criteria}). We denote by $\ell$ the AdS radius, and ${\cal F}=F_{\mu \nu}F^{\mu \nu}$ is the Maxwell invariant, $q$ is a constant related to the total $U(1)$ charge $Q$, and $T$, $S$, $P$, $V$ have their usual meanings in thermodynamics. 

\subsection{(2+1)-Dimensional Black Holes}

Let us first confine our attention to superentropic black holes in three-dimensions (3D). The charged BTZ black hole spacetime is the simplest system that manifests superentropicity. 

For completeness, we recall the relevant thermodynamic quantities for the
 BTZ black hole of mass $M$ and charge $Q$ in the extended phase space, which are obtained from
the line element and corresponding field equations \cite{Mureika2015}:
\begin{align}
	T &= \frac{r_{+}}{2\pi \ell^{2}} - \frac{Q^{2}}{8\pi r_{+}}, \qquad
	S = 4\pi r_{+}, \\[4pt]
	P &= \frac{1}{8\pi \ell^{2}}, \qquad
	V = \pi r_{+}^{2} -\frac{Q^2}{4}\pi\ell^2 \qquad
	\Phi = -\frac{Q}{8} \ln\!\left(\frac{r_{+}}{\ell}\right) \nonumber
\end{align} 
where $r_+$ is the event horizon. 
These are the standard expressions appearing in the extended black hole
thermodynamics of the BTZ family. Using them one directly finds the
isoperimetric ratio
\begin{equation} \label{RII-BTZ-sch1}
{\cal R}_\text{RII} = \sqrt {1 - \frac{{{Q^2}{\ell^2}}}{{4 r_ + ^2}}}  < 1,
\end{equation}
showing the violation of the reverse isoperimetric inequality.

By examining the thermodynamic quantities, the interesting feature for us is the negativity of the adiabatic compressibility, given by 
\begin{equation} \label{AC_chargedBTZ}
{\kappa _S}  =  - \frac{{{Q^2}}}{{32P^2}V} < 0.
\end{equation}
In the $Q \to 0$ limit, the inequality is saturated \cite{Mureika2015} and the adiabatic compressibility vanishes, reflecting a perfectly incompressible system and in agreement with the intuitive picture we presented.

Inclusion of rotation is more subtle. The  original charged rotating BTZ solution \cite{Banados:1992wn,Banados:1992gq} does not satisfy Maxwell's equations; the solution satisfying the full set of Einstein-Maxwell equations was later obtained \cite{Clement:1995zt,Martinez:1999qi} but its thermodynamics has not been fully analyzed.  Curiously, a generalization of (2+1)-dimensional Einstein-Maxwell theory with an additional coupling of  the gauge field to scalar fields \cite{Deshpande:2024vbn} has the original charged rotating BTZ solution as a valid solution  \cite{HeMu2024}.
Thermodynamically $S=\pi r_+/2$ and $V=\pi r^2_+$ for this black hole, so the RII is saturated and ${\kappa _S} =0$. This holds for 
any black hole with non-linear electrodynamics in this more general theory.

However some BTZ-type black holes with non-linear charge coupling have been clained to be counterexamples for the instability conjecture of superentropic black holes \cite{CE2023TorusBH}. Contrary to this claim   it is straightforward to show that $V=\pi r_{+}^{2} -\frac{Q^2}{32 P}$ for these black holes and so  the inequality ${\kappa _S}    < 0$ also holds for this case.

 It is instructive to clarify how our conclusion $\kappa_S<0$ is compatible with the claim made in Ref. \cite{CE2023TorusBH} that a region of parameter space exists in which the specific heats are positive.  
The key point is that positivity of $C_P$ or $C_V$ alone does not guarantee
thermodynamic stability: mechanical stability additionally requires
$\kappa_T \ge \kappa_S \ge 0$.  
For the BTZ and BTZ-like families considered here, the thermodynamic volume $V$ is
a function of the horizon radius whose dependence on pressure is such that
$(\partial V/\partial P)_S>0$ throughout the superentropic regime, leading
inevitably to $\kappa_S<0$ as shown in Eq. (\ref{AC_chargedBTZ}).
Thus even if $C_P>0$ or $C_V>0$ in some parameter region, as reported in
\cite{CE2023TorusBH}, the mechanical stability condition remains violated.
Because the thermodynamic identities (\ref{thermo identities}) link the signs of the specific heats and compressibilities, a negative $\kappa_S$
necessarily precludes simultaneous positivity of both specific heats.  
In this sense there is no contradiction: the positive specific heats reported in
\cite{CE2023TorusBH} correspond to parameter regions in which mechanical
instability persists due to $\kappa_S<0$, so the full set of stability criteria is
never satisfied.  
This distinction underscores why compressibilities, rather than specific heats
alone, provide a sharper diagnostic of superentropic instability.

These results can also be generalized to what are called superentropic charged BTZ-like black holes \cite{DPZS2023}, through a Lagrangian as ${\cal L}_\text{CBTZ-like}={R - 2\Lambda  + {{( - \mathcal{F})}^{(D - 1)/2}}} $, yielding 
\begin{equation}
{\cal R}_\text{RII} = {\left( {1 - \frac{{{2^{(D - 3)/2}}{\ell^2}{Q^{D - 1}}}}{{r_ + ^{D - 1}}}} \right)^{\frac{1}{{D - 1}}}} < 1,
\end{equation}
and 
\begin{equation}
{\kappa _S} =  - \frac{{{{\cal A}_{D - 2}}{2^{(D - 11)/2}}(D - 2){Q^{D - 1}}}}{{\pi {P^2}V}} < 0
\end{equation}
in $D$ dimensions.
In fact the superentropicity originates from the special behavior of Maxwell's electrodynamics in two spatial dimensions and the electromagnetic sector of superentropic charged BTZ-like black holes extends this behavior to higher dimensions. To go beyond Maxwell's theory, we now consider a large class of 3D superentropic black holes by use of couplings of Einstein gravity to nonlinear $U(1)$ theories of electrodynamics (NED) whose weak-field limit, i.e. ${{\cal L}_{{\rm{NED}}}}({\cal F}) =  - {\cal F} + \frac{{{{\cal F}^2}}}{{2{\beta ^2}}} + O\left( {{\beta ^{ - 4}}} \right)$, matches with the weak-field limit of Euler-Heisenberg's theory. The functional form of the Lagrangian densities of NED theories can be normalized in a way that all of them have the same leading order  corrections \cite{DSZ2022}. Generally, for nonlinear $U(1)$ modifications of charged BTZ black holes in the weak-field limit (large $\beta$), one finds ${\left. {\cal R}_\text{RII} \right|_{{\rm{large}}\,\,\beta }} = \sqrt {1 - \frac{{{Q^2}{\ell^2}}}{{r_ + ^2}}}  + {\cal O} \left( {\beta ^{-2}} \right) < 1$ and ${\kappa _S} =   - \frac{{{Q^2}}}{{8{P^2}V}} + {\cal O}\left(\beta^{-4} \right) < 0$. These results can be extended to include the strong-field limit ($\beta \to 0$) where the weak-field expansion is no longer valid. An explicit example is the Born-Infeld charged BTZ metric, which is a solution of the field equations that result from coupling gravity to Born-Infeld Lagrangian ${{\cal L}_{{\rm{BI}}}}(F) = {\beta ^2}\left( {1 - \sqrt {1 + \frac{{2F}}{{{\beta ^2}}}} } \right)$, yielding
\begin{equation}
{\cal R}_\text{RII} = \sqrt {1 - \frac{{{\beta ^2}{\ell^4}}}{{4r_ + ^2}}\left( {\sqrt {1 + \frac{{4{Q^2}}}{{{\beta ^2}{\ell^2}}}}  - 1} \right)}  < 1,
\end{equation}
and
\begin{equation} \label{kappa_3BI}
{\kappa _S} =  - \frac{{{Q^2}\left( {3\sqrt {1 + \frac{{32\pi {Q^2}P}}{{{\beta ^2}}}}  - 1} \right)}}{{8{P^2}V\left( {1 + \frac{{32\pi {Q^2}P}}{{{\beta ^2}}} + \sqrt {1 + \frac{{32\pi {Q^2}P}}{{{\beta ^2}}}} } \right)}} < 0.
\end{equation}

The preceding relations provide strong evidence   that coupling gravity to NED models (also called BI-type electrodynamics, see \cite{DSZ2022}) yields black hole metrics that both   violate the RII and have  negative compressibility.  Eq. (\ref{kappa_3BI}) also implies that the Born-Infeld charged BTZ metric is not a thermodynamically stable superentropic black hole, as claimed in \cite{CE2021BIBTZ}.

The thermodynamic behaviour of generalized exotic BTZ black holes (obtained by a topological coupling of gauge fields to the spin connection \cite{Carlip:1994hq}),  depends crucially on two dimensionless parameters $(\bar{\alpha},\bar{\gamma})$ that appear in the metric function and control the deviation from the standard BTZ case  \cite{Cong2019Mann} ($\bar{\alpha}=1$ corresponds to the ordinary BTZ limit). In particular, one finds that $C_P < 0$ whenever $C_V > 0$ for $\bar{\alpha}< 1/2$.
For exotic black holes in this particular region, we confirmed that $\kappa_{T} < 0$ while $\kappa_{S} > 0$. However, when $\bar{\alpha}> 1/2$, we always observe a negative adiabatic compressibility, $\kappa_{S}<0$. In general, when $\bar{\alpha}> \bar{\gamma}$ superentropic black holes behave mostly standard (having $\kappa_{S}<0$) but in the range $\bar{\alpha} < \bar{\gamma}$ they behave mostly exotic (having $\kappa_{T}<0$) and the usual behaviours of quantities $\kappa_{S}$ and $\kappa_{T}$ are interchanged.

\subsection{Higher Dimensional Superentropic Black Holes}

 Historically, the story of superentropic black holes began with finding a new ultraspinning limit for singly-spinning Kerr-AdS black holes in arbitrary dimensions \cite{UltraSpinningBH2015PRL}.  The value of ${\cal R}_\text{RII}$ for the resultant ultraspinning metric is
 \begin{equation} \label{IR_KerrAdS}
{\cal R}_\text{RII} = {\left( {\frac{{r_ + ^2}}{{r_ + ^2 + {\ell^2}}}} \right)^{\frac{1}{{(D - 1)(D - 2)}}}} < 1.
\end{equation}
which clearly violates the RII conjecture for all regions of parameter space, where
  $r_+$ denotes the event-horizon radius and 
$D$  the spacetime dimension \cite{Cvetic2011, UltraSpinningBH2015PRL, UltraSpinningBH2015JHEP}.
Interestingly, the adiabatic compressibility always takes negative values, given by
\begin{equation} \label{kappaS_KerrAdS}
{{\kappa _S} =   - \frac{{D - 1}}{{\left[ {(D - 1)(D - 4) + 16\pi Pr_ + ^2} \right]P}} < 0}.
\end{equation}
 where  $P$ is the thermodynamic pressure \eqref{press}.

These results are also valid for singly-ultraspinning Kerr-Newman-AdS black holes of charge $Q$.  In four-dimensions we find 
\begin{equation}
{\cal R}_\text{RII} = \left( {\frac{{r_ + ^2}}{{r_ + ^2 + {\ell^2}}}} \right)^{1/6} < 1 
\end{equation}
and 
\begin{equation}
{{\kappa _S} =   - \frac{3}{16\pi P^2r_ + ^2} < 0}
\end{equation}
signifying instability.  It has been argued \cite{CE2023KerrNewmanAdS}  that this particular ultraspinning Kerr--Newman--AdS configuration 
provides a counterexample to the thermodynamic instability conjecture for 
superentropic black holes; our result is clearly contrary to this claim, and shows that this  solution does not constitute a counterexample.

Amongst all superentropic black holes, the ultraspinning limit of doubly-rotating black holes in minimal $5D$ gauged supergravity \cite{UltraSpinningBH2015JHEP} are likely the most complicated superentropic system so far found. In addition to the parameters involved in singly-ultraspinning Kerr-AdS black holes, a charge parameter $Q$ and a second rotation parameter $b$ are present here. Setting $b=0$, one recovers the relevant results of singly-ultraspinning Kerr(-Newman)-AdS black holes in five-dimensions, i.e. the $D \to 5$ limit of relations (\ref{IR_KerrAdS}) and (\ref{kappaS_KerrAdS}). The actual expressions for the entropy and volume are
\begin{eqnarray}\label{thermoQuantities}
S&=& \frac{\mu \pi \left[(b^2+r_+^2)(\ell^2+r_+^2) + b\ell q \right]}{4r_+ \Xi_b}=\frac{A}{4}\,,
\nonumber\\
V &=& \frac{\mu \pi }{12 r_+^2 \ell^2 \Xi_b^2}\Big(\bigl[(b^2+3r_+^2)\ell^2-2b^2r_+^2\bigr](\ell^2+r_+^2)(b^2+r_+^2)  \nonumber\\
   && \qquad + qb\ell\left[(2b^2+3r_+^2)\ell^2 + \ell bq -b^2r_+^2 \right]  \Big) 
 \end{eqnarray}
 where $Q = \frac{\mu \sqrt{3}q}{8 \Xi_b}$ and $\Xi_b = 1-\frac{b^2}{\ell^2}$, with $\mu$ the period of the azimuthal coordinate orthogonal to
 the plane of rotation associated with $b$.
The RII is violated in some regions of the $(Q,b)$ parameter space but not others \cite{UltraSpinningBH2015JHEP}.  
 
\begin{figure}[t]
  \centering
  \includegraphics[width=0.85\linewidth]{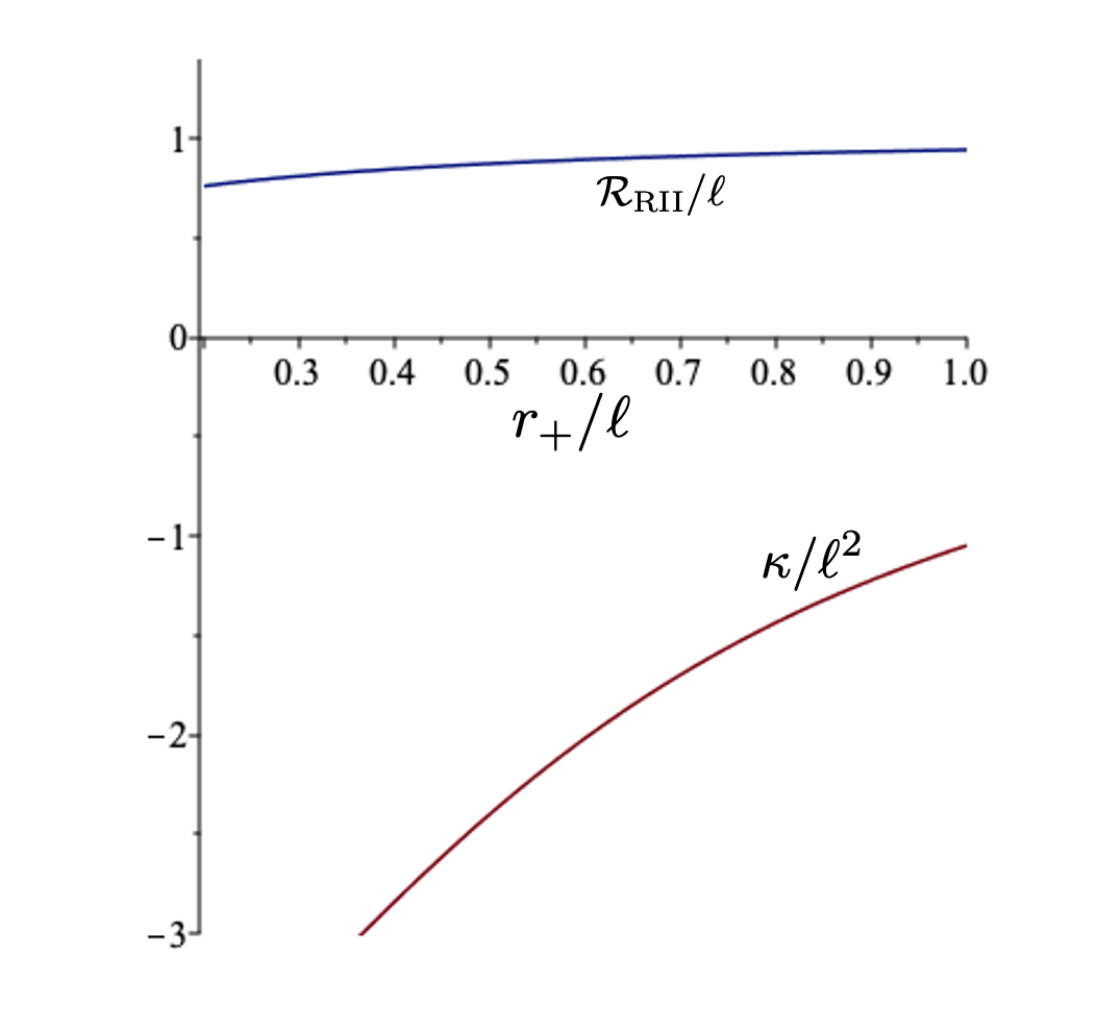}
  \caption{Illustration of the reverse-isoperimetric ratio ${\cal R}_\text{RII}$ (blue)
    and the adiabatic compressibility $\kappa_{S}$  (red) as functions of the
    horizon radius $r_+/\ell$ for   $Q/\ell=0.5$, and $b/\ell^2=0.001$.
    We see that ${\cal R}_\text{RII}<1$ across this parameter range
    whilst $\kappa_{S}$ is negative, emphasizing the mechanical instability
    that accompanies superentropicity in this family.}
  \label{fig:rrii_kappa_example}
\end{figure}

The Taylor expansion about small values of $b$ yields
\begin{equation}
{\left. {{\kappa _S}} \right|_{{\rm{small}}\,b}} =  - \frac{1}{{P + 4\pi {P^2}{r_ + }}} - \sqrt {\frac{\pi }{{3P}}} \frac{{Qb(3 + 4\pi r_ + ^2P)}}{{{{({r_ + } + 4\pi r_ + ^3P)}^2}}} + {\cal O}({b^2}),
\end{equation}
in agreement with the violation of the RII
\begin{equation}
{\left. {\cal R}_\text{RII} \right|_{{\rm{small}}\,b}}  = \left( {\frac{{r_ + ^2}}{{r_ + ^2 + {\ell^2}}}} \right)^\frac{1}{12} - \frac{{blQ}}{{12r_ + ^{11/6}{{\left( {r_ + ^2 + {\ell^2}} \right)}^{13/12}}}} + {\cal O} \left( {{b^3}} \right),
\end{equation}
confirming ${\cal R}_\text{RII}<1$ and $\kappa_{S} < 0$ for small enough $b$.  We have numerically confirmed that 
whenever ${\cal R}_\text{RII} < 1$,  we always found that at least one of   $\kappa_{S}$ or $\kappa_{T}$ are negative whenever $T$ and $V$ are positive.  There are also regions for which  superentropicity corresponds to naked singularities.
To make the connection between superentropicity and mechanical instability explicit, we plot in
figure~\ref{fig:rrii_kappa_example}  the reverse-isoperimetric ratio ${\cal R}_\text{RII}$
and the adiabatic compressibility $\kappa_{S}$ for small $b$ as functions of the horizon radius
for fixed charge. The plot shows that 
${\cal R}_\text{RII}<1$ over the displayed range while $\kappa_{S}$, illustrating directly
that the superentropic regime coincides with negative adiabatic compressibility.

Superentropicity can also enter the physics of black holes in another way, namely  superentropic black holes in 4 spacetime dimensions
with Immirzi hair \cite{Immirzi2021}. These black holes have $M \geq -\ell/4$. For $M=  -\ell/4$ the largest horizon radius is at $r_+ = \ell/2$; for
$M\geq 0$ the horizon  $r_+ \geq \ell$.
The isoperimetric ratio is   \cite{Immirzi2021}
\begin{equation}\label{27}
{\cal R}_\text{RII} = \frac{{{{\left( {3\ell r_ + ^2 - 3{\ell^2}{r_ + } + {\ell^3}} \right)}^{1/3}}}}{{\sqrt {\ell(2{r_ + } - \ell) + {48\alpha }{{({r_ + } - \ell)}^2}/{\ell^2}} }},
\end{equation}
where $r_{+}$ is the horizon radius and $\alpha$ encodes the strength of the 
conformal anomaly correction in the modified field equations, which   characterizes the Jordan frame potential coming from a generalization of the Starobinsky model with the inclusion of a cosmological constant term. 

The isoperimetric ratio is sensitive to the value of the parameter $\alpha$. It has two maxima at $r_+=0$, $r_+ = \ell$ and a local minimum at
$r_+ = \ell^3/(48\alpha)$ provided    $\alpha > \ell^2/48$.  For values of $\alpha$ smaller than this ${\cal R}_\text{RII} >1$ if  $r_+ < \ell$ and the black hole is no  longer superentropic. For  $\alpha \geq \ell^2/24$ both ${\cal R}_\text{RII} \leq 1$ at both local maxima and the black hole is superentropic for all values of $r_+$. For all values of $\alpha$ the black hole is superentropic if $r_+ \geq \ell$.

 Interestingly, for $r_+ > \ell$, we find  
\begin{equation}\label{kappsS-Imm}
{\kappa _S} =  - \frac{{3\sqrt 6  + 4\sqrt \pi  {r_ + }\left( {4\sqrt {6\pi } {r_ + }P - 9\sqrt P } \right)}}{{2\left( {8\sqrt 6 \pi r_ + ^2{P^2} - 12\sqrt {\pi {P^3}} {r_ + } + \sqrt 6 P} \right)}} <0
\end{equation}
where $P=3/8/\pi \ell^2$. For $r_+ = \ell$, we have ${\cal R}_\text{RII} =1$ and $\kappa_{S} = 0$ regardless of the value of $\alpha$.
 The condition $\kappa_{S}<0$  holds for   $r_+ > \ell$, which is the region where ${\cal R}_\text{RII} < 1$, consistent with  with our general 
conjecture.
It is important to emphasize that the expression (\ref{kappsS-Imm}) for the adiabatic
compressibility is physically meaningful only in the regime where the Immirzi
black hole represents a thermodynamic equilibrium state. This requires
non-negative ADM mass, which for these solutions corresponds to $r_+ \ge \ell$.
In the interval $\ell/2 < r_+ < \ell$, the mass is negative and the interpretation
of $M$ as enthalpy in extended black hole thermodynamics breaks down. Although
(\ref{kappsS-Imm}) can be analytically continued into this region and indeed becomes
positive there, this does not signal mechanical stability. Rather, it reflects
the fact that negative-mass Immirzi black holes lie outside the physical domain
of validity of the thermodynamic framework employed here. Restricting to the
physically admissible branch $r_+ \ge \ell$, we find that ${\cal R}_{\rm RII}<1$
always coincides with $\kappa_S<0$, fully consistent with our conjecture.

Motivated by the intuitive picture of superentropicity, by examining a considerable number of superentropic black hole families, we observed that a (either adiabatic or isothermal) compression on superentropic black holes generally causes the volume of the system to increase. So, we \textit{conjecture} that black holes violating RII suffer from negative compressibility (either ${\kappa_S} < 0$ or ${\kappa_T} < 0$ or both), indicating that they cannot represent states of stable or metastable equilibrium. This property together with thermodynamic identities of (\ref{thermo identities}) and adiabatic index (\ref{adiabatic index}) results in violating both the mechanical and thermal stability requirements (\ref{stability criteria}). Finally, we restate the thermodynamic instability conjecture \cite{Johnson2019,Cong2019Mann} in a more stringent way: The mechanical and thermal stability requirements of ${\kappa _T} \geqslant {\kappa _S} \geqslant 0$ and ${C_P} \geqslant {C_V} \geqslant 0$ are never satisfied in superentropic black holes.

We emphasize that in the examples discussed above we have not obtained closed-form expressions for the isothermal compressibility $\kappa_T$. This, however,
does not weaken our conclusions. In all superentropic families for which the
black hole represents a physically admissible equilibrium state, we explicitly
find $\kappa_S<0$. Since mechanical stability requires $\kappa_T \ge \kappa_S
\ge 0$, the negativity of $\kappa_S$ alone is sufficient to rule out mechanical
(and hence thermal) stability, independent of the value of $\kappa_T$.
Moreover, the thermodynamic identity (see (\ref{thermo identities}))
$$
\kappa_T - \kappa_S = \frac{T V \beta_P^2}{C_P}
$$
implies that whenever $T>0$ and $C_P>0$, one necessarily has $\kappa_T \ge
\kappa_S$. Thus a negative $\kappa_S$ already guarantees violation of the
stability conditions. In exotic cases where $\kappa_S>0$, we instead observe
that other stability requirements fail (e.g.\ $\kappa_T<0$ or $C_P<C_V$), so
that the full set of mechanical and thermal stability conditions is never
satisfied in the superentropic regime.

In many superentropic families (such as the ultraspinning Kerr(-Newman)-AdS 
black holes) one finds $\kappa_{S}<0$ directly, while in others 
(e.g.\ generalized exotic BTZ black holes with $\alpha<1/2$) one instead 
encounters $\kappa_{T}<0$ with $\kappa_{S}>0$. 
Thus there are indeed cases where only one of $\kappa_{T}$ or $\kappa_{S}$ is 
negative, but in the superentropic regime we never observe a parameter region 
in which the mechanical stability condition 
$\kappa_{T}\geq \kappa_{S}>0$ is satisfied. 
Via the thermodynamic identities (\ref{thermo identities}) and the adiabatic index relation 
(\ref{adiabatic index}), this also precludes the simultaneous positivity of both specific 
heats with $C_{P}>C_{V}>0$ for superentropic black holes. 
This is the sense in which our conjecture rules out mechanically and 
thermally stable superentropic configurations.
\\

\noindent \textbf{Conjecture 1.} 
Superentropic black holes (i.e. those with $R_{\text{RII}}<1$) never satisfy the mechanical and thermal stability requirements 
\(\kappa_T > \kappa_S > 0\) and \(C_P > C_V > 0\). 
In other words, violation of the reverse isoperimetric inequality necessarily implies thermodynamic instability.

\section{There is no asymptotically flat limit of superentropic black holes}

Asymptotically flat limits of AdS black holes are generally obtained by taking the limit $\ell \to \infty$ (or equivalently $\Lambda \to 0$)\footnote{The only exception occurs in $D = 7$ dimensions where additional subtleties arise.}. A fundamental question arises: do superentropic black holes possess well-defined asymptotically flat limits? This is, in fact, not unexpected: the ultraspinning construction itself requires the rotation parameter to saturate the AdS radius, $a \to \ell$. 
Consequently, sending $\ell \to \infty$ destroys the ultraspinning limit, so a smooth asymptotically flat limit cannot exist. 
This point was already emphasized in Ref. \cite{UltraSpinningBH2015JHEP}.
Through systematic analysis of various superentropic black hole families, we find that such limits are generically problematic.

Consider the charged BTZ black hole with isoperimetric ratio $R_{RII} = \sqrt{1 - Q^2 \ell^2/r_+^2} < 1$.  In the flat limit $\ell \to \infty$, while maintaining finite physical quantities, the charge parameter must scale as $Q \sim \ell^{-1}$, which either eliminates the superentropic character ($R_{RII} \to 1$) or leads to divergent thermodynamic quantities. We note that the flat limit formally yields a metric without horizon that is not static.

Similarly, for ultraspinning Kerr-AdS black holes with $R_{RII} = (r_+^2/(r_+^2 + \ell^2))^{1/((D-1)(D-2))} < 1$, the flat limit requires careful treatment of the rotation parameter. The constraint for maintaining superentropicity conflicts with the requirement of finite angular momentum in flat space, leading to either loss of superentropic character as $R_{RII} \to 1$, divergent angular momentum or other pathological behavior, or violation of geometric constraints for event horizon existence.

For nonlinear electrodynamics models (Born-Infeld, etc.), the situation is even more restrictive. The Born-Infeld parameter $\beta$ must satisfy specific scaling relations in the flat limit to preserve superentropicity, but these typically lead to either vanishing electromagnetic effects or divergent field strengths.

\textbf{Conjecture 2:} There exists no superentropic black hole spacetime that has a smooth limit to an asymptotically flat background.

The physical interpretation is profound: if a superentropic black hole could smoothly approach flat space, it would necessarily violate the Penrose isoperimetric inequality (PII) in the limit, since $R_{RII} < 1$ would persist while the thermodynamic volume approaches the geometric volume. This would create a naked singularity, violating cosmic censorship. Thus, superentropicity serves as a natural  `censor'  preventing the formation of naked singularities through the flat space limit.

This conjecture has important implications for the AdS/CFT correspondence, suggesting that superentropic states in the bulk correspond to fundamentally non-normalizable or pathological states in the boundary theory. Furthermore, it provides a new perspective on the cosmic censorship hypothesis: the RII conjecture may act as a geometric constraint ensuring that physically reasonable black holes emerge only when appropriate causal structure is maintained.

The absence of flat limits for superentropic black holes also suggests that extended thermodynamics with $\Lambda$ as a thermodynamic variable is not merely a mathematical convenience, but captures essential physics that cannot be reduced to the $\Lambda \to 0$ case.

In nature, generally, any physical system has an adiabatic index greater than 1. The well-known examples are monatomic, diatomic, and triatomic gases. However, for superentropic black holes we quite generally observe $\gamma < 1$. The only exceptions are those ones that could have states with negative values for all of the specific heats as well as the compressibilities. In such cases, that we call super-unstable states, we also  observe $\gamma > 1$.

For any thermodynamic system in nature, there is a speed of sound associated with the adiabatic compressibility, defined by $v_s^2 = {\left. {\frac{{\partial P}}{{\partial \rho }}} \right|_S} = \frac{1}{{1 + \rho {\kappa _S}}}$ where the derivative is taken isentropically (at constant $S$). All known examples in nature as well as AdS black holes confirm the result ${v_s} \in [0,1]$ \cite{Dolan2011}. From the negativity of ${\kappa _S}$, it is now inferred that superentropic spacetimes could always exceed the upper bound, ${v_s} > 1$, for a certain range of the parameters.

\section{Flat Riemannian background}

AdS spacetimes naturally appear in the landscape of superstring theory and also supergravity. Besides this, the main motivation for studying AdS black holes and their thermodynamic properties comes from the AdS/CFT correspondence which establishes a connection between certain kinds of gauge field theories in $D$-dimensions and AdS gravity models in one higher dimension. Further new insights have been gained due to the discovery of the thermodynamic definition of black hole volume \cite{Review2017CQG}.
Remarkably, the definition of thermodynamic volume in AdS is useful in accessing a smooth limit to that of asymptotically flat black holes \cite{Cvetic2011,Dolan2011}, which basically is different from the naive geometric definition, proposed in \cite{Parikh2006}. The perspective that we take here is to generalize the applications of extended black hole thermodynamics in AdS to the real-world physics of a four-dimensional flat Riemannian background, more than before: we claim one can gain insights from extended AdS black hole thermodynamics to  a flat Riemannian background to find novel isoperimetric inequalities. This can open novel paths towards the study of cosmic censorship   and naked singularities.

Thanks to the thermodynamic volume, isoperimetric inequalities in terms of thermodynamic and geometric volumes have been computed for black holes with AdS \cite{Cvetic2011} and dS \cite{Dolan2013deSitterIso} asymptotes. In AdS the RII conjecture (\ref{RII}) is satisfied;  otherwise the black hole is superentropic \cite{UltraSpinningBH2015PRL} and, as we have explicitly shown, suffers from mechanical and thermal instabilities. In a dS background, the RII conjecture has been also shown to remain valid for the volumes and horizons associated with the black hole and the cosmological horizons \cite{Dolan2013deSitterIso}. However, the black hole volume in between the event and the cosmological horizons is equal to the naive geometric definition (see the definition in Eq. (\ref{naive_vol}) below), and for such cases, the CII (\ref{II_Euclidean}) holds instead \cite{Dolan2013deSitterIso}. Here, we are primarily interested in such isoperimetric inequalities (if any) for asymptotically flat black holes. 
In cosmology, current observations indicate that the Universe has nearly flat spatial sections (with curvature parameter $k\simeq 0$). 
This, however, is distinct from the notion of asymptotic flatness in general relativity, which refers to the metric approaching 
the Minkowski form at spatial infinity. 
Nevertheless, since the cosmological constant is extremely small on astrophysical scales, black holes relevant for astrophysics are 
well approximated by asymptotically flat solutions.

Before proceeding further, let us clarify what we mean by the naive geometric volume ($V'$), presented in \cite{Parikh2006}; that is a geometrically simple, slicing-invariant definition of volume for asymptotically flat stationary spacetimes, defined by
\begin{equation} \label{naive_vol}
V' = \frac{{d{V_D}(t)}}{{dt}} = \int_{{r_0}}^{{r_ + }} {dr} \int {{d^{D - 2}}x\sqrt { - {g_{(D)}}} },
\end{equation}
where $d{V_D}(t) = \int_t^{t + dt} {dt'} \int {dr} \int {{d^{D - 2}}x\sqrt { - {g_{(D)}}} }$ and $g_{(D)}$ is the determinant of the $D$-dimensional spacetime. This definition, despite its role in dS space for the volume between the horizons, does not have a universal characteristic in AdS. Here \(r_0\) denotes the inner radial cutoff of the integration region. 
For ordinary static black holes without inner horizons it is natural to set \(r_0=0\), 
so that $V'$ measures the proper volume from the center of the black hole up to the outer event horizon at $r=r_+$. 
In spacetimes with an inner horizon, one may instead take $r_0$ to be the inner horizon radius.  

It is natural to inquire whether analogous versions of RII (\ref{RII}) or CII (\ref{II_Euclidean}) hold in the limit   $\Lambda\to 0$ (equivalently $\ell \to\infty$), which we refer to as the  flat limit. In an asymptotically flat Riemannian background, we generally observe $V \ge V'$ (obviously, this fails for superentropic AdS black holes, which, as we already pointed out,  have no asymptotically flat limit).

The thermodynamic volume for asymptotically flat black holes is defined by the smooth $\Lambda\to 0$ limit of the AdS construction. 
Starting from the generalized Komar identity (\ref{KomarInt}) and the extended first law with $P=-\Lambda/(8\pi G_N)$, we write
\begin{equation}
V_{\text{flat}}
\;\equiv\;
\left(\frac{\partial M}{\partial P}\right)_{S,J,Q,\ldots}\Bigg|_{P=0}
\;=\;
-\,8\pi G_N\,
\left(\frac{\partial M}{\partial \Lambda}\right)_{S,J,Q,\ldots}\Bigg|_{\Lambda=0}.
\label{eq:Vflat-def}
\end{equation}
For the black hole families considered here $M(S,P,\ldots)$ is analytic in $P$ near $P=0$, so the limit exists and yields a finite result. 
This limiting prescription is equivalent to introducing an auxiliary cosmological constant, computing the enthalpy $M(S,P,\ldots)$ in AdS via the Komar/Killing--potential expression for $V$ (Eq. (\ref{volume-thermo})), and finally sending $P\to 0$. 
If additional matter fields are present, the corresponding work terms are held fixed in the derivative at fixed $(S,J,Q,\ldots)$; the definition \eqref{eq:Vflat-def} still applies provided $M(S,P,\ldots)$ is smooth at $P=0$.

We have considered conventional static black holes, Myers-Perry black holes in any dimension (the $\Lambda \to 0$ limit of General Kerr-AdS metrics), the $\Lambda \to 0$ limits of AdS black holes in gauged supergravity theories, and also black rings. In all cases we have found that the RII conjecture  
\begin{equation} \label{RII_flat}
{\left( {\frac{{(D - 1)V}}{{{{\cal A}_{D - 2}}}}} \right)^{\frac{1}{{D - 1}}}} \ge {\left( {\frac{A}{{{{\cal A}_{D - 2}}}}} \right)^{\frac{1}{{D - 2}}}},
\end{equation}
remains valid, 
where $V = {\left. {\frac{{\partial {M_{{\rm{(AdS)}}}}}}{{\partial P}}} \right|_{P \to 0}}$.

 This statement has been explicitly verified for standard asymptotically flat
black hole families obtained as smooth $\Lambda\to0$ limits of AdS solutions.
As a representative example, consider the Kerr black hole in $D=4$ dimensions.
The ADM mass $M$, horizon area $A$, and thermodynamic volume $V$ obtained from
the flat limit of the AdS enthalpy are given by
\begin{equation}
M = \frac{r_+^2 + a^2}{2 r_+}, \qquad
A = 4\pi (r_+^2 + a^2),
\end{equation}
and
\begin{equation}
V = \frac{4\pi r_+}{3} \left( r_+^2 + a^2 \right)
\left( 1 + \frac{a^2}{2 r_+^2} \right),
\end{equation}
where $r_+$ is the outer horizon radius and $a$ is the rotation parameter.
The volume $V$ coincides with the thermodynamic volume defined via
\begin{equation}
V = \left.\frac{\partial M_{\text{AdS}}}{\partial P}\right|_{S,J}
\end{equation}
after taking the limit $P\to0$.\\
The reverse isoperimetric ratio in flat space then becomes
\begin{equation}
{\cal R}_{\text{RII}}
=
\left(
\frac{3V}{4\pi}
\right)^{1/3}
\left(
\frac{4\pi}{A}
\right)^{1/2}
=
\left(
1 + \frac{a^2}{2 r_+^2}
\right)^{1/3}
\ge 1,
\end{equation}
with equality attained only in the non-rotating limit $a\to0$.
Hence the flat-space reverse isoperimetric inequality is satisfied
for Kerr black holes, and is strictly stronger than the Penrose
isoperimetric inequality in the rotating case.   We have carried out
analogous explicit checks for Schwarzschild,
Reissner--Nordstr\"{o}m, Kerr--Newman (in $D=4$), and Myers--Perry black
holes in $D=5$, all of which satisfy Eq.~(\ref{RII_flat}) when the thermodynamic
volume is defined by the smooth AdS-to-flat limit.

A critical question emerges regarding the universal applicability of different volume definitions in asymptotically flat isoperimetric inequalities. While the thermodynamic volume $V$ (obtained via the generalized Komar integral) exhibits universal characteristics in AdS backgrounds, the naive geometric volume $V'$ defined by Eq.~(\ref{naive_vol}) fails to demonstrate such universality in curved spacetimes~\cite{Cvetic2011}.

In AdS space, Cveti\v{c} et al.~\cite{Cvetic2011} demonstrated through charged solutions of gauged supergravity that the geometric volume $V'$ does not satisfy universal isoperimetric relations. This failure stems from the geometric volume's inability to capture the full thermodynamic structure encoded in the generalized first law. The key question is whether this distinction persists in the flat space limit.

Our analysis reveals that for physically reasonable asymptotically flat black holes -- those admitting smooth limits from AdS configurations that respect ${\cal R}_{RII} \geq 1$ -- both volume definitions yield consistent isoperimetric hierarchies. Specifically, we find the thermodynamic volume systematically exceeds the geometric volume, $V \geq V'$, leading to the inequality chain:

\begin{equation} \label{VAV'_flat}
{\left( {\frac{{(D - 1)V}}{{{{\cal A}_{D - 2}}}}} \right)^{\frac{1}{{D - 1}}}} \ge {\left( {\frac{A}{{{{\cal A}_{D - 2}}}}} \right)^{\frac{1}{{D - 2}}}} \ge {\left( {\frac{{(D - 1)V'}}{{{{\cal A}_{D - 2}}}}} \right)^{\frac{1}{{D - 1}}}},
\end{equation}
where the middle inequality is a standard isoperimetric-type
statement for the geometric volume $V'$.
Since our $V'$ is defined by the spacetime determinant as in Eq. (\ref{naive_vol}),
rather than as the Euclidean volume of a region in $\mathbb{R}^{D-1}$,
this step is not automatic and must be verified for the relevant
families of asymptotically flat black holes.

 To make this explicit, consider the Kerr black hole in $D=4$.
Using the definition (\ref{naive_vol}) with $r_0=0$ and the Boyer--Lindquist form
of the metric, one has $\sqrt{-g^{(4)}}=\rho^2\sin\theta$ with
$\rho^2=r^2+a^2\cos^2\theta$.
Carrying out the angular integrals gives
\begin{eqnarray}
V'_{\rm Kerr}
&=&
\int_0^{r_+} dr \int d\Omega_2\, \sqrt{-g^{(4)}}
=
4\pi \int_0^{r_+} dr\left(r^2+\frac{a^2}{3}\right)\nonumber\\
&=&
\frac{4\pi r_+}{3}\left(r_+^2+\frac{a^2}{3}\right).
\end{eqnarray}
The horizon area is $A=4\pi(r_+^2+a^2)$, and therefore
\begin{equation}
\left(\frac{A}{\mathcal{A}_2}\right)^{\!1/2}
=
\sqrt{r_+^2+a^2}
\ge
\left(\frac{3V'_{\rm Kerr}}{\mathcal{A}_2}\right)^{\!1/3}
=
\left(r_+^3+\frac{a^2 r_+}{3}\right)^{\!1/3},
\end{equation}
which confirms the middle inequality in Eq. (\ref{VAV'_flat}) for Kerr.
 Had we set $r_0=r_{-}$,   the resultant $V'$ would be even smaller and the inequality would still hold.
Equality is recovered only in the non-rotating limit $a\to 0$,
where $V'$ reduces to the usual Euclidean ball volume $4\pi r_+^3/3$.
The final inequality in (\ref{VAV'_flat}) follows from the general observation that the thermodynamic volume is never smaller 
than the naive geometric volume, $V \geq V'$ (see  \cite{Cvetic2011,Parikh2006}).  

More generally, we have checked this geometric inequality for the
standard asymptotically flat families that admit smooth AdS embeddings, including Schwarzschild, Reissner--Nordstr\"om (where the inequality is saturated), Kerr,  and Kerr--Newman metrics in $D=4$, and Myers--Perry in $D=5$ (and in the equal-spin subsectors in higher $D$), and found no violations.
We emphasize that we are not claiming a general theorem for arbitrary stationary spacetimes and arbitrary slicings; rather, Eq. (\ref{VAV'_flat}) is supported by these explicit black hole families together with the general inequality $V\ge V'$.
Since both sides enter with the same monotonic power $1/(D-1)$, the inequality directly translates to the 
dimensionless quotients shown above. 
Physically, this reflects the fact that $V$ encodes additional gravitational binding-energy contributions 
absent in $V'$, so that the effective thermodynamic volume systematically exceeds the purely geometric measure.

To avoid confusion, let us emphasize what is and is not being claimed here. 
For asymptotically flat black holes we established the hierarchy (\ref{VAV'_flat}), namely that the thermodynamic volume 
is always greater than or equal to the naive geometric volume, $V \geq V'$, and hence the flat-space version 
of the RII lies between the Penrose inequality and the purely geometric bound. 
We are not claiming a general analytic theorem that RII violation in the flat case necessarily implies negative compressibility, 
but we have found this to hold in all explicit asymptotically flat black hole families we have examined.
Specifically, these include:
(i) Schwarzschild and Reissner--Nordstr\"om black holes,
(ii) Kerr and Kerr--Newman black holes in $D=4$,
(iii) Myers--Perry black holes in $D=5$ with one or more rotation
parameters,
(iv) the $\Lambda\to0$ limits of Kerr--AdS and Kerr--Newman--AdS solutions,
and
(v) black rings, for which the Penrose and reverse isoperimetric
inequalities hold but the thermo-volumetric inequality fails, consistent
with their known dynamical instabilities.
In all these cases, whenever ${\cal R}_{\rm RII}<1$ occurs, we find that at least
one of the mechanical stability conditions ($\kappa_S<0$ or
$\kappa_T<0$) is violated. We emphasize that our claim is based on explicit analytic calculations in $D=4,5$ and
numerical checks over the admissible parameter space for the families
listed, rather than on a general proof valid for arbitrary stationary
spacetimes.

This hierarchy reflects a fundamental physical principle: the thermodynamic volume encodes gravitational binding energy effects absent in the naive geometric definition. For rotating black holes, this difference becomes particularly pronounced, with $V = V'(1 + a^2/((D-2)r_+^2))$ for Kerr-type solutions.

Importantly, the charged gauged supergravity solutions that violate geometric volume universality in AdS~\cite{Cvetic2011} do not admit smooth flat space limits due to their superentropic character (as established in our Conjecture 2). This suggests that the apparent universality of the geometric volume in flat space is a consequence of the restricted class of physically realizable asymptotically flat black holes, rather than an intrinsic property of the geometric definition.

The thermodynamic volume thus emerges as the physically preferred quantity for isoperimetric inequalities across all asymptotic backgrounds, maintaining consistency between AdS and flat space descriptions while properly encoding the gravitational thermodynamics structure.

As a physically important example, for all Myers-Perry black holes in any spacetime dimension, we find 
$${{\cal R}_{{\rm{CII}}}} = {\left( {\prod\limits_{i = 1}^N {\frac{{{r_+^2}}}{{{r_+^2} + a_i^2}}} } \right)^{\frac{1}{{(D - 1)(D - 2)}}}} \le 1$$ 
whereas for the reverse isoperimetric quotient we find
\begin{align}
{{\cal R}_{{\rm{RII}}}} &= {\left( {1 + \frac{1}{{(D - 2){r_+^2}}}\sum\limits_{i = 1}^N {a_i^2} } \right)^{\frac{1}{(D - 1)}}}{\left[ {\prod\limits_{i = 1}^N {\left( {1 + \frac{{a_i^2}}{{{r_+^2}}}} \right)} } \right]^{ - \frac{1}{(D - 1)(D - 2)}}} \nonumber \\
&= {\left( {1 + \frac{1}{{(D - 2){r_+^2}}}\sum\limits_{i = 1}^N {a_i^2} } \right)^{\frac{1}{(D - 1)}}} {{\cal R}_{{\rm{CII}}}} 
\end{align}
and following what has already been proved for general Kerr-AdS black holes \cite{Cvetic2011}, one finds ${\cal R}_\text{RII} \ge 1$.  These results establish the endpoint inequalities
${\cal R}_{\rm RII}\ge 1$ and ${\cal R}_{\rm CII}\le 1$.
They do not, by themselves, imply the intermediate bound involving the
horizon area in Eq.(\ref{VAV'_flat}), which must be checked separately. For the Myers--Perry family,  the middle inequality
\begin{equation}
\left(\frac{A}{\mathcal A_{D-2}}\right)^{\!1/(D-2)}
\ge
\left(\frac{(D-1)V'}{\mathcal A_{D-2}}\right)^{\!1/(D-1)}
\end{equation}
we have verified explicitly.
In particular, for equal-spin Myers--Perry black holes and for all
single-spin cases in $D=4,5$, direct evaluation of the naive geometric
volume $V'$ defined in Eq. (\ref{naive_vol}) shows that the area-based bound is satisfied, with equality attained only in the static limit.
As noted above, we find that (\ref{VAV'_flat})  holds in a broad variety of explicit examples;   we do not claim a general proof of the middle inequality
for arbitrary stationary spacetimes or arbitrary slicings.
 
Now we have two isoperimetric inequalities in the flat Riemannian  background: the PII (\ref{Penrose inequality}) and the RII (\ref{RII}). Is there any relationship between these two? Going back to the PII (\ref{Penrose inequality}), we can rewrite it in terms of the dimensionless quantity of Penrose isoperimetric quotient ${\cal R}_\text{PII}$ as
\begin{equation}\label{35}
{\cal R}_\text{PII} \equiv {\left( {\frac{{16\pi E}}{{(D - 2){{\cal A}_{D - 2}}}}} \right)^{\frac{1}{{D - 3}}}}{\left( {\frac{{{{\cal A}_{D - 2}}}}{A}} \right)^{\frac{1}{{D - 2}}}} \ge 1.
\end{equation}
 In comparison with the RII (\ref{RII}), we ask whether  the following two quantities
\begin{equation}
 {\left( {\frac{{16\pi E}}{{(D - 2){{\cal A}_{D - 2}}}}} \right)^{\frac{1}{{D - 3}}}} \mathop  \leftrightarrow \limits^?  {\left( {\frac{{(D - 1)V}}{{{{\cal A}_{D - 2}}}}} \right)^{\frac{1}{{D - 1}}}}
\end{equation}
have some   characteristic relationship.  We find that these two quantities are equal  for Schwarzschild black holes in any dimension. However in other cases, such as Reissner-Nordstr\"{o}m black holes,  Kerr(-Newman) black holes in $D=4$ and $D=5$ dimensions, as well as static black holes in Einstein gravity coupled with Born-Infeld-type nonlinear electrodynamics and Lovelock gravity theories, we find that the Penrose isoperimetric quotient ${\cal R}_\text{PII}$ is greater than that of  RII (\ref{RII}), ${\cal R}_\text{PII} \ge {\cal R}_\text{RII}$. This suggests the string of inequalities  
\begin{equation} \label{IsoInequalities}
{\left( {\frac{{16\pi E}}{{(D - 2){{\cal A}_{D - 2}}}}} \right)^{\frac{1}{{D - 3}}}} \ge {\left( {\frac{{(D - 1)V}}{{{{\cal A}_{D - 2}}}}} \right)^{\frac{1}{{D - 1}}}} \ge {\left( {\frac{A}{{{{\cal A}_{D - 2}}}}} \right)^{\frac{1}{{D - 2}}}},
\end{equation}
meaning that the asymptotically flat limit of the RII remarkably provides two new isoperimetric inequalities that are stronger than the PII. 

The left inequality, which includes the mass and   thermodynamic volume (the thermo-volumetric inequality), is new in black hole physics.  We emphasize that this inequality is derived entirely within the framework of black hole thermodynamics, where both the ADM mass $M$ and the thermodynamic volume $V$ are related by a Smarr relation and originate from gravitational dynamics.
At present, we do not claim a general theorem excluding its violation in ordinary thermodynamic systems; rather, we are not aware of any known non-gravitational system for which an inequality of the form
$M \gtrsim V^{1/(D-1)}$ is universally satisfied. The reason black holes are distinguished in this respect is that their
thermodynamic variables are not independent: the mass, entropy, and
thermodynamic volume are linked by the Smarr relation, reflecting the underlying scale invariance of the gravitational field equations.
In contrast, for ordinary laboratory systems the internal energy and volume can be varied independently, and no universal lower bound on the energy in terms of the volume is known. For this reason, the thermo-volumetric inequality should be viewed as a
gravitational isoperimetric bound rather than as a general constraint on arbitrary thermodynamic systems.

The right (the RII) is the reverse of the CII (\ref{II_Euclidean}) in which the geometric volume in Euclidean space is promoted to the thermodynamic volume in flat Riemannian space. We remark that the implication RII $\Rightarrow$ PII in flat space follows directly once the left inequality in (\ref{IsoInequalities}),
\[
{\cal R}_{\text{PII}} \;\ge\; {\cal R}_{\text{RII}},
\]
is established.  We have verified this analytically for Schwarzschild and Kerr black holes
in $D=4,5$, where the horizon geometry and thermodynamic volume depend on a single length scale (or a single rotation parameter).
In higher dimensions, generic rotating black holes (e.g. Myers--Perry) carry multiple independent rotation parameters, and both the horizon area and the thermodynamic volume become nontrivial symmetric polynomials in these parameters.
As a result, the difference ${\cal R}_{\rm PII}-{\cal R}_{\rm RII}$ does not factorize in a way that admits a simple analytic sign determination. For $D\ge6$, we therefore employed numerical checks over the physically allowed parameter space (existence of horizon, $M>0$, $T>0$), and found that while both the Penrose and reverse isoperimetric inequalities continue to hold individually, the relative ordering between them is no longer universal once multiple spins are present. Thus the restriction of analytic proofs to $D=4,5$ reflects increasing algebraic complexity rather than a conceptual limitation of the inequalities themselves.

In Eqs. (\ref{RII_flat})-(\ref{IsoInequalities}) we use $V$ exclusively for the thermodynamic volume, defined as 
$V \equiv \left(\frac{\partial M}{\partial P}\right)_{S,J,Q,\ldots}\vert_{\Lambda=0}$ (equivalently via the Komar-Killing potential in~(\ref{volume-thermo})), while $V'$ denotes the naive geometric volume introduced in Eq. (\ref{naive_vol}). 
Equation (\ref{IsoInequalities}) involves $V$ (not $V'$). Replacing $V$ by $V'$ in (\ref{IsoInequalities}) would contradict the hierarchy in (\ref{VAV'_flat}), and is not intended.
To be clear, our claim is that in the asymptotically flat case the reverse isoperimetric inequality (RII) 
implies the Penrose isoperimetric inequality (PII), but not conversely. 
That is, whenever the flat-space RII is satisfied, the PII automatically follows, 
so that PII is a strictly weaker bound. 
For a Schwarzschild black hole both inequalities coincide and are saturated, 
while for rotating solutions such as Kerr one finds  ${\cal R}_{\text{PII}} \;\ge\; {\cal R}_{\text{RII}}$
so that the RII provides the stronger constraint. 
This is the precise sense in which we state later in result 2 that PII is weaker than RII.
We have confirmed these for a large class of black hole systems in Einstein's theory and beyond in $D=4$ and $D=5$ dimensions. However, for black rings ($D \ge 5$) we observe that both the RII and PII hold but the thermo-volumetric inequality (left inequality in \eqref{IsoInequalities})
is violated; however these black objects are known to be unstable for the whole parameter space. 

A highly interesting, non-trivial case to examine is the Kerr(-Newman) family of black holes since it can discriminate between the thermodynamic volume and the geometric volume. The thermodynamic volume and the ADM mass in $D$-dimensions from the $\Lambda \to 0$ limit of generalized Komar integral (\ref{KomarInt}) reads
\begin{equation}\label{eq49}
M = {\left. {{M_{{\rm{AdS}}}}} \right|_{\Lambda  \to 0}}= \frac{{{{\cal A}_{D - 2}}(D - 2)}}{{16\pi }}\left( {r_ + ^2 + {a^2}} \right)r_ + ^{D - 5},
\end{equation}
\begin{equation}\label{eq50}
V = {\left. {\frac{{\partial {M_{{\rm{(AdS)}}}}}}{{\partial P}}} \right|_{P \to 0}}=V' \left[ {1 + \frac{a^2}{{(D - 2)r_ + ^2}}} \right],
\end{equation}
where $A = {{\cal A}_{D - 2}}\left( {r_ + ^2 + {a^2}} \right)r_ + ^{D - 4}$ and $V' =\frac{{A{r_ + }}}{{D - 1}}$ is the naive geometric definition of the volume.  Equations (\ref{eq49}) and (\ref{eq50}) satisfy all the isoperimetric inequalities
presented in Eq.(\ref{IsoInequalities}) for Kerr black holes in $D=4,5$.
The Kerr--Newman case is included only in $D=4$, where it likewise
respects Eq.(\ref{IsoInequalities}). For $D \ge 6$, we observe a violation of the thermo-volumetric inequality
\eqref{IsoInequalities}
for rotating vacuum solutions of Myers--Perry type with multiple
independent rotation parameters, although both the RII and the PII
remain valid individually. Interestingly, for Myers--Perry black holes with only a single nonzero
rotation parameter, in any spacetime dimension, we did not find any
violation of the thermo-volumetric inequality. We emphasize that statements concerning $D\ge6$ refer exclusively to
Myers--Perry black holes; no Kerr--Newman analog exists in those
dimensions. Thus the apparent tension is resolved: violations of the thermo-volumetric inequality arise only when multiple angular momenta are simultaneously present, whereas single-spin Myers--Perry black holes satisfy all inequalities in Eq. (\ref{IsoInequalities}).

We summarize our three key results thus far.
\begin{enumerate}
\item  The RII conjecture holds for physically reasonable black holes in flat and AdS spacetimes. Superentropicity in any background corresponds to either thermodynamic instabilities or no physically acceptable black hole solutions (such as those with negative temperature) or naked singularities.

\item  For asymptotically flat Schwarzschild, Reissner--Nordstr\"om, Kerr, and Kerr--Newman black holes   in $D=4$, and singly-rotating Myers--Perry black holes in $D\ge 5$, 
we find that
\begin{equation}\label{ineqchain}
{\cal R}_{\rm PII} \ge {\cal R}_{\rm RII} \ge 1,
\end{equation}
implying  that the Penrose isoperimetric inequality (PII) is strictly weaker than the reverse isoperimetric inequality (RII) in the asymptotically flat limit. 
Equality is attained only for Schwarzschild black holes.
We conjecture that this hierarchy continues to hold more generally in higher dimensions, 
although for $D\geq 6$ with multiple rotation parameters the thermo-volumetric inequality (left step of the chain) can fail, 
even though both PII and RII remain valid individually. 
Establishing the circumstances in which \eqref{ineqchain} holds  remains an open problem.

\item  We already pointed out that an asymptotically flat limit of superentropic black holes is not possible in any case. This means it is impossible to respect the classical form of the isoperimetric inequality (\ref{II_Euclidean}) in terms of thermodynamic volume while having a curvature singularity (with closed trapped region) covered by an event horizon in Riemannian flat backgrounds. Otherwise, the PII is likely violated and a naked singularity would be the result.  

\end{enumerate}

We also speculate that black holes respecting both the PII and RII conjectures but violating  the thermo-volumetric inequality 
(left inequality in \eqref{IsoInequalities}) have some type of   instability. This merits further investigation. Note that since generally the thermodynamic definition of volume is equal or greater to the naive geometric one, there is a weaker isoperimetric inequality  $V \ge V'$. 
In this context, one can also formulate an inequality involving the ADM mass and the geometric volume that is weaker than the Penrose isoperimetric inequality. This inequality has already been studied in the literature as the volumetric Penrose inequality, proven for conformally flat spacetimes~\cite{Schwartz2011}.

For clarity, we summarize why the claims in Refs. \cite{CE2021BIBTZ, Immirzi2021, CE2023KerrNewmanAdS, CE2023TorusBH} do not contradict our instability conjecture. 
In each case the apparent discrepancy originates from either (i) using a volume other than the thermodynamic one, 
(ii) working in an ensemble where only heat capacities are tested while compressibilities are ignored, 
or (iii) including parameter regions that violate basic physicality constraints (existence of horizon, $T>0$, $M>0$).

Stability cannot be inferred from $C_P>0$ alone. Mechanical stability also requires $\kappa_T\ge\kappa_S\ge 0$. 
Using the Komar/Killing-potential definition of thermodynamic volume  \eqref{volume-thermo} we find for the Born-Infeld BTZ family  \eqref{kappa_3BI}, across the physically allowed domain ($T>0$, horizon present) that  even where $C_P>0$, mechanical stability fails due to $\kappa_S<0$.
Conclusions drawn from $C_P$ alone miss this obstruction \cite{CE2021BIBTZ}.

The superentropic sector arises only in the ultraspinning limit; for the genuine superentropic branch we obtain 
${\cal R}_{\text{RII}}<1$ and Eq. (\ref{kappaS_KerrAdS}) (and its $D{=}4$ specialization). 
Parameter ranges highlighted as stable in non-ultraspinning regimes are not superentropic; 
in the ultraspinning/superentropic regime $\kappa_S$ is strictly negative, so the mechanical (and hence thermal) stability 
conditions cannot be met \cite{Immirzi2021}.

 In particular, in Refs.~\cite{CE2021BIBTZ,Immirzi2021}
the isoperimetric ratio is evaluated either using   geometric volume (rather than the thermodynamic (Komar) volume), or using a nonstandard normalization of $\mathcal A_{D-2}$ appropriate to planar or toroidal horizon identifications.
In such cases the resulting quantity differs from the reverse
isoperimetric ratio defined in Eq. (\ref{RII}), and comparisons with the RII must be interpreted with care. When the thermodynamic volume defined via the Killing potential and the
correct horizon normalization are used consistently, the apparent
counterexamples reported in these references disappear.
For planar/toroidal horizons the unit-area factor and identifications enter nontrivially; using the Komar-based $V$ and the 
correct horizon normalization restores the RII assessment, and the putative stable superentropic region disappears once 
$\kappa_{S,T}$ are computed \cite{CE2023KerrNewmanAdS}. 

 Thermodynamic variables are sometimes mixed between different ensembles, and the volume employed is not the Komar/Killing-potential (thermodynamic) volume.
This occurs, for example, in Refs.\cite{CE2021BIBTZ,Immirzi2021,CE2023CG},
where stability is assessed using heat capacities alone or using a
geometric volume rather than the thermodynamic volume.
In particular, Ref. \cite{CE2021BIBTZ} analyzes stability in
ensembles that fix the geometric volume or omit the pressure--volume sector entirely, while Ref. \cite{Immirzi2021} employes volume definitions that do not arise from the extended first law with mass interpreted as enthalpy. When the analysis is reformulated using a consistent extended
thermodynamic framework, with mass as enthalpy, pressure $P=-\Lambda/8\pi$,
thermodynamic volume from Eq. \eqref{volume-thermo}, and the full set of stability conditions $\kappa_T\ge\kappa_S\ge0$ and $C_P\ge C_V\ge0$, the apparent counterexamples disappear.

Imposing a consistent extended-thermodynamics ensemble (mass as enthalpy, $V$ from Eq.~(\ref{volume-thermo})) and 
enforcing $T>0$, horizon existence, and $M>0$, we find regions identified as stable either fail basic physicality 
or exhibit $\kappa_S<0$ (hence fail mechanical stability \cite{CE2023TorusBH}).

In summary, the cited references either (a) assess stability using only heat capacities without testing the compressibilities 
required by Eq.~(\ref{stability criteria}), (b) use  geometric volume instead of  thermodynamic volume (thereby altering $R_{\text{RII}}$), or 
(c) include unphysical parameter regions. 
Reanalyzing within a consistent extended-thermodynamic framework (mass as enthalpy, $V$ from Eq.~(\ref{volume-thermo}), and the 
full set of stability criteria) removes the apparent counterexamples.

\section{Asymptotically flat Lovelock black holes with $P=0$.}

In certain Lovelock branches one has $\Lambda=0$ and hence $P=0$ identically (see e.g.~\cite{arXiv:2212.08087}). 
In this case the extended first law carries no $V\,dP$ term if $\Lambda$ is not varied. 
Nevertheless, the thermodynamic volume in the flat limit is well-defined by introducing an auxiliary cosmological constant, 
computing the AdS enthalpy $M(S,P,\ldots)$, and then taking the smooth limit $P\to 0$:
\begin{eqnarray}
V_{\text{flat}}
\;&\equiv&\;
\left(\frac{\partial M}{\partial P}\right)_{S,J,Q,\ldots}\Bigg|_{P=0}\nonumber\\
\;&=&\;-\,8\pi G_N\left(\frac{\partial M}{\partial \Lambda}\right)_{S,J,Q,\ldots}\Bigg|_{\Lambda=0},
\label{eq:Vflat-def-Lovelock}
\end{eqnarray}
which is the Lovelock analogue of Eq.~\eqref{eq:Vflat-def}. 
With $V\equiv V_{\text{flat}}$ the flat-space reverse isoperimetric inequality retains the form
\begin{equation}
\left(\frac{(D-1)V}{A_{D-2}}\right)^{\!1/(D-1)} \;\ge\;
\left(\frac{A}{A_{D-2}}\right)^{\!1/(D-2)},
\label{eq:RII-flat-Lovelock}
\end{equation}
 and the inequality chain in Eq. (\ref{VAV'_flat}) (the thermo-volumetric bound together with the PII) continues to apply for all
explicitly known asymptotically flat Lovelock black hole
solutions for which the thermodynamic volume can be
unambiguously defined. We have explicitly checked this for static asymptotically flat Lovelock black holes  in the cases where the thermodynamic volume obtained via the limiting procedure coincides with the natural geometric volume and the reverse isoperimetric inequality is saturated.
For rotating Lovelock black holes, fully general asymptotically
flat solutions are not presently known, and no claim of an
explicit check is made here.
In such cases, the validity of Eq. \eqref{VAV'_flat} should be regarded as
a conjectural extension, motivated by continuity from the AdS
case and by the absence of any known counterexamples.

Now, we summarize implications for our conjectures.

For $P\equiv 0$, compressibilities $\kappa_{S,T}$ are not directly defined within the $P$-fixed ensemble. 
The appropriate interpretation is via the limiting ensemble: embed the solution in a family with small $P>0$, 
evaluate $\kappa_{S,T}$, and then take $P\to 0^+$. 
If a Lovelock solution were to satisfy $R_{\mathrm{RII}}<1$ in such a family, Conjecture 1 implies $\kappa_S<0$ or $\kappa_T<0$ 
in a punctured neighborhood of $P=0$, so no stable equilibrium would exist in the limit either. 
We are not aware of physically acceptable asymptotically flat Lovelock black holes with $R_{\mathrm{RII}}<1$.

For static asymptotically flat Lovelock black holes with $P{=}0$, we find that physically reasonable solutions satisfy \eqref{eq:RII-flat-Lovelock} (i.e.\ $R_{\mathrm{RII}}\ge 1$ with $V=V_{\text{flat}}$).
Using $V=V_{\text{flat}}$, we expect the string of inequalities in Eq. (\ref{VAV'_flat}) to hold . 

In summary, asymptotically flat Lovelock black holes with $P{=}0$ should satisfy the flat-space RII and the PII, 
and are not superentropic when $V$ is defined by \eqref{eq:Vflat-def-Lovelock}. 
If a counterexample with $R_{\mathrm{RII}}<1$ were found, the limiting-ensemble reasoning above would trigger 
the instability mechanism of Conjecture~1 near $P=0$.

\section{Concluding remarks and future outlook}

 We have investigated the physical significance of the reverse isoperimetric inequality (RII) in black hole thermodynamics and its relation to the Penrose isoperimetric inequality (PII).
Using extended black hole thermodynamics, we showed that violations of the RII are not merely geometric curiosities but have direct physical consequences.

Our central result is that superentropic black holes
(${\cal R}_{\rm RII}<1$) never satisfy the combined mechanical and thermal stability conditions
$\kappa_T \ge \kappa_S \ge 0$ and $C_P \ge C_V \ge 0$.
Through explicit calculations across a wide class of examples in
Einstein gravity and beyond, we found that superentropicity is always accompanied by negative compressibility or other thermodynamic pathologies, including negative temperature, naked singularities, or the absence of a physically admissible equilibrium state.
This leads us to a strengthened instability conjecture: violation of the reverse isoperimetric inequality necessarily implies thermodynamic instability.

We further argued that superentropic black holes do not admit smooth asymptotically flat limits.
Whenever such a limit is forced, either the superentropic character is lost or physical pathologies arise.
This observation motivates the conjecture that asymptotically flat black holes must satisfy the flat-space version of the RII, thereby preventing violations of the Penrose inequality and protecting cosmic censorship.

Taking the $\Lambda\to0$ limit of extended black hole thermodynamics, we derived a hierarchy of isoperimetric inequalities in asymptotically flat spacetimes.
For all explicit families examined in four and five dimensions, including rotating black holes 
in the cases considered here 
we found
\begin{equation}
{\cal R}_{\rm PII} \ge {\cal R}_{\rm RII} \ge 1,
\end{equation}
showing that the Penrose inequality is strictly weaker than the reverse isoperimetric inequality.
The left-hand inequality defines a new thermo-volumetric bound, relating the ADM mass to the thermodynamic volume.
While this bound holds for all four-dimensional examples studied, it can fail for certain higher-dimensional rotating solutions, indicating that its validity is sensitive to dimensionality and solution space.

Our results highlight the thermodynamic volume, defined via the generalized Komar construction, as the physically preferred notion of volume for black holes.
Unlike purely geometric definitions, it yields consistent isoperimetric relations in AdS, de Sitter, and asymptotically flat backgrounds, and admits a smooth flat-space limit.
From this perspective, the RII emerges as a unifying principle linking black hole geometry, thermodynamic stability, and cosmic censorship.

Several open questions remain.
A general proof relating RII violation to negative compressibility would be of particular interest, as would a deeper understanding of the dynamical endpoint of superentropic instabilities.
More broadly, our results suggest that extended black hole thermodynamics provides a powerful framework for uncovering universal constraints on gravitational systems, both in AdS and in physically relevant flat-space settings.

\section*{Acknowledgements}

This work was supported in part by the Natural Sciences and Engineering Research Council of Canada.

\end{document}